\documentclass[prd,amsfonts,twocolumn,nofootinbib,showpacs]{revtex4}
\RequirePackage{graphicx,  amsmath}
\begin{document}

\title{The hybrid curvaton}
\author{Konstantinos Dimopoulos}
\affiliation{Consortium for Fundamental Physics,
Physics Department, Lancaster University, Lancaster LA1 4YB, UK}
\author{Kazunori Kohri}
\affiliation{Cosmophysics group, Theory Center, IPNS, KEK,
and The Graduate University for Advanced Study (Sokendai),
Tsukuba 305-0801, Japan}
\author{Tomohiro Matsuda}
\affiliation{Physics Department, Lancaster University,  
Lancaster LA1 4YB, UK, and
 Laboratory of Physics, Saitama Institute of Technology,
Fukaya, Saitama 369-0293, Japan}

\begin{abstract}
\hspace*{\parindent}
We study both oscillating and inflating curvaton scenarios when the curvaton
mechanism is caused by a hybrid potential.
The source of the curvature perturbation is 
the inhomogeneous phase transition that causes the modulation 
of the onset of the oscillation.
For the supergravity-motivated curvaton there is a possibility of finding
 natural coincidence of the energy density that is needed to affect
non-Gaussianity in the curvaton scenario.   
\end{abstract}
\maketitle

\section{Introduction}
The primordial perturbation in the inflationary epoch is widely believed
to be the primary origin of the formation of the large-scale
structure in the Universe~\cite{Lyth-book}.
The generation of the curvature perturbation $\zeta$ in the original
inflation scenario is due to the inflaton perturbation $\delta \phi$
that exits the horizon during inflation.
The curvature perturbation $\zeta$ in the original single-field
inflation scenario is generated from the inflaton perturbation $\delta 
\phi$, which remains constant after the horizon exit.
In the curvaton scenario, $\zeta$ is instead 
generated after inflation from the isocurvature perturbation of a curvaton
field~\cite{Curvaton-paper} (for a recent review see Ref.~
\cite{anupam-review}). 

\begin{figure}[htb]
\centering
\includegraphics[width=1.0\columnwidth]{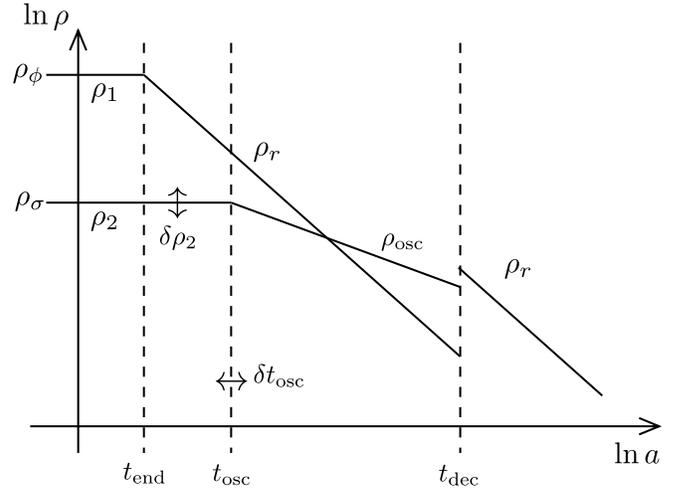}
\caption{Evolution of the energy density of the curvaton and the
 background radiation. The graph shows the evolution of the
 ``oscillating curvaton'', which dominates the energy density of
 the Universe during oscillation. The crossing ($\rho_{\mathrm{osc}}
=\rho_{r}$)
 does not happen if $r_\mathrm{dec}<1$. As is discussed in the text, $r_\mathrm{dec}$ is
 determined by the model parameters.}  
\label{fig:usual}
\end{figure}

In this paper we consider a hybrid potential for the curvaton scenario
 and study both the original (oscillating)~\cite{Curvaton-paper} and
 the inflating curvaton~\cite{infla-curv}.~\footnote{See also
 Ref.~\cite{Furuuchi:2011wa} for another application of the inflating
 curvaton.} The hybrid potential is very important in cosmological
 model building and a vast variety of models have been
 considered~\cite{locked-inf, vast-varietyHyb}.  This paper presents
 the first attempt in choosing the hybrid potential $V(\varphi,
 \sigma)$ for the curvaton mechanism.  In this paper we try to keep
 the discussion as simple as possible.  We introduce three parameters
 $\alpha$, $\beta$ and $\gamma$ to measure the model dependence of the
 scenario, and avoid highly model-dependent arguments.  Since the
 non-trivial evolution after the waterfall~\cite{Lyth-hybrid} may ruin
 the simplicity of the model, we assume matter-like evolution
 ($\rho_\sigma\propto a^{-3}$) that starts just after the phase
 transition.  The phase transition is called ``trigger'' in this paper
 because the phase transition triggers the oscillation of the
 waterfall field $\sigma$.  A crucial difference from the usual
 curvaton is that the source of the perturbation is not the
 perturbation of the oscillating (waterfall) field (because $\delta
 \sigma=0$ for large scale) but the perturbation of the slow-rolling
 field ($\delta \varphi\ne 0$)~\cite{Matsuda:2008-9}, which is not the
 inflaton ($\phi$) of the primordial inflation but it is the field
 which triggers the phase transition.  The characteristics of this
 scenario are shown in Fig.~\ref{fig:usual}.

Let us first consider the basic curvaton scenario.
The initial perturbation of the curvaton $\sigma$ is given by  
\begin{equation}
\left.\frac{\delta \rho_\sigma}{\rho_\sigma}\right|_\mathrm{osc}\simeq
 2\frac{\delta 
 \sigma}{\sigma}.
\end{equation}
This corresponds to the perturbation $\delta \rho_2$ in
Fig.\ref{fig:usual}.   If the oscillation starts  at $m_\sigma^2\simeq
H^2$, $t_{\mathrm{osc}}$ (in the figure) is exactly determined by
the curvaton mass $m_\sigma$. Then  there is no perturbation which
corresponds to $\delta t_\mathrm{osc}$.  Obviously, both $\delta
\rho_2$ and $\delta t_\mathrm{osc}$ can source the curvaton
perturbation, but in the basic scenario the source perturbation is
almost entirely coming from $\delta \rho_2$.

The total curvature perturbation during the curvaton oscillation is
given by
\begin{equation}
\zeta=(1-f)\zeta_r + f\zeta_\sigma,
\end{equation}
where we introduced the ratio $f\equiv 3\rho_\sigma/(4\rho_r+3\rho_\sigma)$.
Here $\zeta_r\equiv \delta \rho_r/4\rho_r$ is negligible.
To calculate the curvaton perturbation, what we need is the quantity
$\zeta_\sigma$ at the beginning of the 
oscillation because $\zeta_\sigma$, defined by 
\begin{equation}
\zeta_\sigma \equiv -H\frac{\delta \rho_\sigma}{\dot{\rho}_\sigma}
=\frac{1}{3}\frac{\delta \rho_\sigma}{\rho_\sigma},
\end{equation}
becomes constant during the (quasi)sinusoidal oscillation.

The new idea in this paper is that $\zeta_\sigma$ {\em is dominated by
the perturbation} $\delta t_\mathrm{osc}$. 
In the basic (non-hybrid) scenario, the perturbation caused by  $\delta
 t_\mathrm{osc}$ does not 
 dominate the total curvature perturbation.\footnote{Modulated
 interaction may cause the curvature perturbation to be dominated by $\delta
 t_\mathrm{osc}$~\cite{Matsuda:2008-9, Matsuda:2007tr}. Such modulated
 (inhomogeneous) interaction is not considered in this paper.}
The reason is as follows.
The source of $\delta t_\mathrm{osc}$ in the basic curvaton scenario 
is $\delta m_\sigma$, and 
$m_\sigma$ is not a constant when the potential is not quadratic.
Assuming that the sinusoidal oscillation starts at
$\eta=1$, the perturbation at the beginning of the
oscillation is given by
\begin{equation}
\delta t_\mathrm{osc}\simeq -
\frac{\delta \eta}{\dot{\eta}} 
\end{equation}
Here we consider
\begin{eqnarray}
\delta \eta &\simeq&\frac{\partial \eta}{\partial\sigma}\delta \sigma 
\\
\label{doteta}
\dot{\eta}&\simeq& \frac{\partial\eta}{\partial\sigma}\dot{\sigma}
+ \frac{\partial\eta}{\partial H}\dot{H}
\simeq -2\eta \frac{\dot{H}}{H^2} H 
=4\eta H,
\end{eqnarray}
where, in contrast to the inflation phase, 
$\dot{H}/H^2=-2$ in the radiation dominated Universe, which dominates
the evolution of $\eta$.
Assuming a polynomial function $\eta \propto \sigma^p$, we find
$\partial \eta/\partial \sigma = p\eta/\sigma$ and
\begin{equation}
\delta t_\mathrm{osc}\simeq \frac{p}{4H}\frac{\delta \sigma}{\sigma},
\end{equation}
where the quadratic potential gives $p=0$, which always leads to
$\delta t_\mathrm{osc}=0$.
For $p\ne 0$, we find 
\begin{equation}
\left.\frac{\delta \rho_{\sigma}}{\rho_\sigma}\right|_{\delta t_\mathrm{osc}}
\simeq  
\frac{\dot{\rho}_{\sigma} \times \delta t_\mathrm{osc}}{\rho_\sigma}
=-3 H\delta t_\mathrm{osc}\simeq
\frac{3p}{4}\frac{\delta \sigma}{\sigma}.
\end{equation}
The above perturbation is comparable to the perturbation sourced by
$\delta \rho_2$ in the basic curvaton scenario. 
Therefore, this perturbation does not dominate the curvature
perturbation, but may change its non-Gaussianity. 

As the result, the deviation from the quadratic potential ($p\ne 0$)
may cause $\delta t_\mathrm{osc}\ne 0$, but this perturbation cannot dominate
over the total curvature perturbation. 
In the original curvaton scenario, the above dynamics 
has been included in the function  
$g(\sigma_*)$ introduced in Ref.~\cite{Lyth-g-func}.
(See also Ref.\cite{Dimopoulos:2003ss,
Kawasaki:2011pd})

\section{Non-Inflating Hybrid curvaton}
To find a significant enhancement of the curvaton perturbation, consider
a hybrid potential that is given by
\begin{equation}
V=\frac{\lambda}4\left(M^2-\sigma^2\right)^2 +\frac{g^2}{2} \varphi^2 \sigma^2 +\frac{1}{2}m_\varphi^2 \varphi^2.
\end{equation}
The potential during slow-roll is $V_0\simeq
\frac{\lambda}{4}M^4$. 

The hybrid potential usually leads to generation of cosmic
strings at the phase 
transition ($\sigma=0\rightarrow \sigma\ne 0$), but in this curvaton
scenario the strings do not lead  to a serious cosmological problem
because the typical energy scale is much lower than
the cosmological bound.

Suppose that the ``rolling field'' ($\varphi$) rolls from far away
($\varphi_\mathrm{ini}\gg\varphi_c$) and the sinusoidal
oscillation of the ``waterfall field'' ($\sigma$) starts suddenly
after the phase transition.
In this model, $t_\mathrm{osc}$ is determined by the ``rolling field''
reaching the waterfall at $\varphi(t_\mathrm{osc})=\varphi_c$.,
where \mbox{$\varphi_c\equiv(\sqrt\lambda/g)M$} is the critical value which 
triggers the phase transition.
Therefore, the perturbation $\delta t_\mathrm{osc}$~\cite{2nd-wands} is
given by 
\begin{equation}
\delta t_\mathrm{osc} \simeq 
\frac{\delta \varphi}{\dot{\varphi}},
\end{equation}
where $\dot{\varphi}$ denotes the velocity in the
radiation-dominated Universe evaluated at the waterfall.  
On the other hand, the initial value of the curvaton oscillation is given by
$\rho_\sigma(t_\mathrm{osc})=V_0=\frac{\lambda}{4}M^4$, which means that the conventional source of the
perturbation ($\delta \rho_2$ in Fig.~\ref{fig:usual}) 
vanishes in this model. 
For simplicity, assume that the energy density of the curvaton oscillation
evolves as $\rho_\sigma\propto a^{-3}$ just after the phase transition.

In this scenario, the perturbation of the ``rolling field''
($\delta \varphi\ne 0$) is converted into $\delta t_\mathrm{osc}$ to cause 
the perturbation $\delta \rho_\sigma$.
In this model, the transfer of the perturbation
$\delta \varphi \rightarrow \delta t_\mathrm{osc}\rightarrow \delta \rho_\sigma$
sources the curvaton perturbation
$\zeta_\sigma\equiv \delta \rho_\sigma/3\rho_\sigma$.

The modulation of the phase transition ($\delta t_\mathrm{osc}\simeq
\delta \varphi /\dot{\varphi}$) causes the 
perturbation of the curvaton density;
\begin{equation}
\label{eq_convert}
\frac{\delta \rho_\sigma}{\rho_\sigma} \simeq
 \frac{\dot{\rho}_\sigma \delta t_\mathrm{osc}
}{\rho_\sigma} 
\simeq -3H 
\frac{\delta \varphi}{\dot{\varphi}},
\end{equation}
where $\dot{\rho}_\sigma=-3H\rho_\sigma$ is assumed for the oscillation.
If $\varphi$ rolls slowly before the phase transition, there is a 
significant enhancement of the curvaton perturbation.
Note that, in the above equation, $\dot{\varphi}$ is not the 
conventional slow-roll velocity which is usually given by
$\dot{\varphi}_s\equiv -V_\varphi/3H$. 
This field has a significant acceleration $\ddot{\varphi}\ne 0$ due to 
$\dot{H}/{H^2}=-2$ in the radiation-dominated Universe.
Moreover, the typical expansion given by
$\delta \varphi = \dot{\varphi}\delta t +
 \frac{1}{2}\ddot{\varphi}(\delta t)^2+...$ shows that the relation
 between $\delta t$ and $\delta \varphi$ is obviously non-linear.
This kind of non-linearity does not appear in the case of the standard 
quasi-de Sitter inflation (with $\varphi$ the inflaton), since 
$\ddot{\varphi}\simeq 0$ is usually assumed 
for $\varphi$ to undergo slow-roll.
However,  $\ddot{\varphi}\simeq 0$ is not true in the
radiation-dominated background.

For later convenience, we consider the approximation;
\begin{equation}
\dot{\varphi}(t_\mathrm{osc})\simeq \alpha\dot{\varphi}_s\equiv
- \alpha\frac{V_\varphi}{3H_\mathrm{osc}}, 
\end{equation}
where $H_\mathrm{osc}^2\equiv \rho_r/3M_p^2> V_0/3M_p^2$ denotes the
Hubble parameter at the phase transition.
Here $M_p$ is the reduced Planck mass. 
The deviation from slow-roll is measured by 
$\alpha \ne 1$. 
We assume $\alpha \sim {\cal O}(1)$ to avoid extreme
situations.\footnote{Although it depends on  situation, the
assumption $\alpha \sim {\cal O}(1)$ may hold even if the field $\varphi$ 
started oscillating near the waterfall. 
Here the breakdown of slow-roll occurs first, then fast-roll may
follow. There could be a breakdown of fast-roll if $\beta>1$
($\beta$ is defined in Eq.~(\ref{beta})), where
$\varphi$ 
starts oscillating but we assume the oscillation soon stops at the
waterfall. 
Assuming that fast-roll breaks down at $\varphi=\varphi_b$,
$\varphi$ gains the kinetic energy 
$K_\Delta<\frac{1}{2}m_\varphi^2\varphi_b^2$.
Comparing $K_\Delta$ with the (possible) slow-roll kinetic energy at the 
waterfall;
$K_s\simeq \left[\frac{m_\varphi^2\varphi_c}{3H}\right]^2$, we find
approximately $\alpha < \varphi_b/\varphi_c$.
Note that in Eq.~(\ref{eq_convert}) the conventional slow-roll is not assumed.}
In some extreme models the non-Gaussianity could be affected, but in the
present scenario the effect is at most ${\cal O}(1)$.
In the radiation-dominated Universe and for the quadratic potential,
$\alpha=3/5$ has been derived in Ref.~\cite{Kawasaki:2011pd}. 
The opposite ($\rho_r <V_0$) corresponds to the
inflating curvaton scenario. 
An intermediate scenario where $H_\mathrm{osc}^2\sim V_0/3M_p^2$ is also
discussed in this paper.

Considering the oscillation of $\sigma$ as in  the basic
curvaton scenario, the density perturbation of the curvaton is 
\begin{equation}
\zeta_\sigma\equiv -H\frac{\delta \rho_\sigma}{\dot{\rho}_\sigma}
=\frac{1}{3}\frac{\delta \rho_\sigma}{\rho_\sigma}.
\end{equation}
From Eq.(\ref{eq_convert}), we find 
\begin{eqnarray}
\zeta_\sigma&\simeq& -\left(H \frac{\delta
                     \varphi}{\dot{\varphi}}\right)_\mathrm{osc}\\
&=&\left(\frac{1}{\alpha} \frac{3H^2}{m_\varphi^2}
    \frac{\delta 
\varphi}{\varphi}\right)_\mathrm{osc}\\
&\equiv& \left(\frac{1}{\alpha \eta} \frac{\delta
                                             \varphi}{\varphi}\right)_\mathrm{osc},
\end{eqnarray}
where $\eta\equiv m_\varphi^2/3H^2$ denotes the slow-roll parameter of
the rolling field $\varphi$. 
For later convenience, we introduce a new parameter $\beta$ that is
defined by
\begin{equation}
\beta \equiv  \frac{m_\varphi^2}{m_0^2},
\label{beta}
\end{equation}
where $m_0^2 \equiv \frac{V_0}{M_p^2}$ is the possible soft-breaking
mass that usually appears in supergravity when the F-term is
given by $|F|\sim V_0^{1/2}$. It is evident that $\beta\sim 1$ is motivated by 
supergravity, and $\beta\not\sim 1$ measures the deviation from the naive
setup.\footnote{D-term inflation
may avoid $\beta\sim 1$~\cite{D-terminf, D-termmatsuda}.
$\beta >1$ is inevitable if the usual supersymmetry breaking sector
(which is not related to the curvaton potential) gives the mass term with
$\beta>1$. In both cases $\beta$ is highly model-dependent.} 
$\eta \ll 1$ and $\beta\sim 1$ are possible at the same time if the
Universe is dominated by the radiation~\cite{Lyth-moroi}.

\subsection{Slow-roll and slow-decay}

Let us consider the spectrum
${\cal P}^{1/2}_{\delta \varphi} \simeq H_1/2\pi$ for the slow-rolling field,
where $H_1$ is the Hubble parameter during the primordial inflation.
Defining the energy ratio at the beginning of the
$\sigma$ oscillation as 
$$r_\mathrm{osc} \equiv
\frac{V_0}{\rho_r(t_\mathrm{osc})}=\frac{\lambda M^4}{12M_p^2 
H_\mathrm{osc}^2}\ll 1\,,$$ 
we find at $t_\mathrm{osc}$;
\begin{equation}
\label{eta-beta}
\eta = \beta r_\mathrm{osc}\;.
\end{equation}
Again, $\eta \ll \beta$ is possible when $r_{\rm osc}\ll 1$.
The spectrum of the curvaton perturbation is given by
\begin{equation}
{\cal P}^{1/2}_{\zeta_\sigma} \simeq \frac{1}{
 \alpha\beta r_{\rm osc}}
 \left(\frac{H_1/2\pi}{\varphi}\right)_\mathrm{osc}
\simeq \frac{g}{2\pi \alpha\beta r_{\rm osc}\sqrt{\lambda}}\frac{H_1}{M}\,.
\end{equation}
Here the evolution of $\delta \varphi$ is neglected because of
the slow-roll assumption.
Defining the gravitational decay constant as 
$\Gamma_g\simeq (\sqrt{\lambda}M)^3/M_p^2$, 
and considering the actual decay rate 
$\Gamma_\sigma\equiv\xi \Gamma_g\ge\Gamma_g$, where $\xi (\gtrsim 1)$
measures the deviation from the gravitational decay,
 we find that the Hubble parameter at
the decay is given by
\begin{equation}
H_\mathrm{dec} \simeq \Gamma_\sigma \simeq \xi\frac{\lambda^{3/2}M^3}{M_p^2},
\end{equation}
which leads to the density ratio
\begin{equation}
r_\mathrm{dec}\equiv\left.\frac{\rho_\sigma}{\rho}\right|_\mathrm{dec}
=r_\mathrm{osc}
\times \left(
\frac{H_\mathrm{osc}}{H_\mathrm{dec}}\right)^{1/2}
\end{equation}
where the evolution is possible until $r_{\rm osc}=1$.
Here the value of $r_\mathrm{osc}<1$ (at the beginning of the oscillation) is
determined by the initial condition.
Although the situation depends on the choice of the model, $\xi\gg 1$
may cause significant interaction 
between $\varphi$ and the radiation, which could alter the
potential. 
In this sense, the potential used in the above argument is the effective
potential approximated at the point very close to the waterfall and it
includes all these corrections.  

The spectral index is $n-1\simeq -2\epsilon_1+2\eta$ and the
non-Gaussianity is  $f_{NL}=\frac45 r_\mathrm{dec}^{-1}$ for 
$r_\mathrm{dec}\ll1$, as in the
usual curvaton scenario.
Here $\epsilon_1\equiv \dot{H}/H^2$ and $\eta\equiv m_\varphi^2/3H^2$
are defined at the horizon exit.

\subsection{Slow-roll and fast-decay}
As a second example, we consider the fast decay scenario.
Again, we find the spectrum of the perturbation given by
\begin{equation}
{\cal P}^{1/2}_{\zeta_\sigma} \simeq \frac{1}{
 \alpha\beta r_\mathrm{osc}}
 \left(\frac{{\cal P}^{1/2}_{\delta\varphi}}{\varphi}\right)_\mathrm{osc}
\simeq \frac{g}{2\pi \alpha\beta r_\mathrm{osc}\sqrt{\lambda}}\frac{H_1}{M}.
\end{equation}
Here the fast-decay assumption leads to $H_\mathrm{dec}\sim H_\mathrm{osc}$.
To quantify the ratio, we introduce a parameter defined by
\begin{equation}
P_d \equiv \frac{H_\mathrm{dec}}{H_\mathrm{osc}}\le1,
\end{equation}
where $P_d=1$ corresponds to the instant decay.
We find 
\begin{equation}
r_\mathrm{dec}\equiv\left.\frac{\rho_\sigma}{\rho}\right|_\mathrm{dec}
=\left.\frac{\rho_\sigma}{\rho_r}\right|_\mathrm{osc}\times \left(
\frac{H_\mathrm{osc}}{H_\mathrm{dec}}\right)^{1/2}
= \frac{r_{\mathrm{osc}}}{\sqrt{P_d}}.
\end{equation}
As in the usual curvaton scenario, the phase transition at
$r_\mathrm{dec}\ll 1$ leads to enhanced non-Gaussianity.
The model predicts that, whenever $r_\mathrm{osc} \lesssim 1$ is natural
in the hybrid curvaton scenario, the instant decay may affect
non-Gaussianity. 
This is not mandatory in the slow-roll scenario, but could be
mandatory if slow-roll breaks down before the waterfall.
We are going to consider this possibility in the next section, although
a more exact calculation requires numerical study. 

\subsection{Breakdown of slow-roll and fast decay}
Since the radiation energy density is decreasing while $m_\varphi$ 
is constant, the natural idea is that $\eta \sim 1$ 
triggers the significant variation of $\varphi$ and eventually
it causes the phase transition.
Because the contribution of $\varphi$ to the scalar 
potential is small (i.e, $m_\varphi^2\varphi^2 \ll V_0$), 
the kinetic energy of $\varphi$ is not significant even if the
slow-roll is violated before the waterfall.\footnote{The opposite limit
leads to oscillations of the same type as in 
locked inflation~\cite{locked-inf}, which is not considered in this paper.}
To avoid an extreme situation, our modest assumption in this section is
$\alpha \sim 1$.
On the other hand, the perturbation $\delta \varphi$ leaves
horizon during the primordial inflation, when the slow-roll parameters
of $\varphi$ are supposed to be much smaller than unity.
We thus find that the spectral index is small, $n-1\ll 1$.

The problem in the slow-roll scenario is that, since the variation of
$\varphi$ (i.e, $\Delta \varphi\equiv 
\varphi(t_\mathrm{ini})-\varphi_c$) is negligible,
we always have to assume careful tuning of the initial condition
$\varphi(t_\mathrm{ini})\simeq 
\varphi_c$.
This tuning is relaxed if slow-roll breaks down
just before the phase transition.
Moreover, since $\beta \sim1 $ is justified in supergravity, 
there is the {\em important prediction} that $r_\mathrm{osc}\lesssim 1$ in the 
hybrid curvaton model.

To look into more details, consider the evolution in
the radiation-dominated Universe.
This immediately leads to the equation of
motion
\begin{equation}
\ddot{\varphi}+3H(t) \dot{\varphi}+(cH(t)^2 + m_\varphi^2)\varphi=0,
\end{equation}
where the Hubble parameter evolves as
\begin{equation}
H(t)=\frac{1}{2t}.
\end{equation}
Before fast-roll, we assume $m_\varphi^2 \ll cH(t)^2$.
Substituting $dt=4td\tau/3$, we find
for the radiation dominated Universe~\cite{Dimopoulos:2003ss}   
\begin{equation}
\frac{d^2 \varphi}{d\tau^2}+ \frac{2}{3}\frac{d \varphi}{d\tau}
+\frac{3c}{9}\varphi \simeq 0,
\end{equation}
which has an oscillating solution when
\begin{equation}
c \ge \frac{1}{4}.
\end{equation}
According to Ref.~\cite{Lyth-moroi}, there can be a cancellation that
leads to $c\ll 1$ during radiation domination.
Therefore, a conceivable set-up of the model is 
$c\ll 1/4$ with fast-roll starting when
$m_\varphi^2=H(t_\mathrm{fast})^2/4$.
At this moment ($t=t_\mathrm{fast}$), we find 
\begin{equation}
\rho_r =12 m_\varphi^2 M_p^2=12\beta V_0\;.
\end{equation}

To capture the evolution after the breaking of slow-roll,
we introduce $\gamma
\equiv t_\mathrm{osc}/t_\mathrm{fast}\ge 1$, which is determined by 
the evolution thereafter.
$\gamma =1$ means that the $\sigma$-oscillation starts immediately after
the breaking of slow-roll. Note that a
non-trivial evolution of the waterfall field can be described by
$\gamma\gg 1$, which we will not consider. There could be a small-scale
perturbation of $\delta \sigma$, but this is not the topic of this paper. 

For $\gamma^2<12\beta$, the $\sigma$-oscillation starts in the
radiation-dominated Universe.
This means that $r_\mathrm{osc}$ is given by
\begin{equation}
\label{beta-ro}
r_\mathrm{osc} =\frac{\gamma^2}{12\beta}<1.
\end{equation}
Assuming fast decay ($P_d \sim 1$) and $\beta\sim 1$, the
slow-roll breakdown just before the waterfall may produce
naturally the coincidence $r_\mathrm{dec}\lesssim 1$.
Therefore, the scenario gives rise to $r_\mathrm{dec}\lesssim 1$, which
can affect non-Gaussianity.

One may suspect that a large initial value ($\varphi\sim M_p$) can easily
cause a long-time rolling that leads to the curvaton domination 
before the phase transition. This speculation is true.
In this respect, the scenario more or less depends on the initial condition.
$\varphi(t_\mathrm{ini})\gg \varphi_c$ with $\beta\sim 1$ may lead
to the opposite scenario, in which 
domination by the curvaton starts {\em before} the phase transition.
This scenario of the inflating curvaton~\cite{infla-curv} will be
discussed in the next section. 

In the above scenario with $\beta\sim 1$, long-time
oscillation of the waterfall field is not needed for the  curvaton domination. 
The waterfall field $\sigma$ can decay after a few oscillations,
{\em in contrast to the standard curvaton scenario}, in 
which a small decay rate is needed for the curvaton domination.

The spectrum of the curvature perturbation is obtained in
the same way as in the previous (slow-roll) scenario; 
\begin{equation}
{\cal P}^{1/2}_{\zeta_\sigma} \simeq \frac{1}{
 \alpha\beta r_\mathrm{osc}}
 \left(\frac{{\cal P}^{1/2}_{\delta \sigma}}{\varphi}\right)_\mathrm{osc}\simeq \frac{1}{
 \alpha\beta r_\mathrm{osc}}
 \left(\frac{{\cal P}^{1/2}_{\delta \sigma}}{\varphi}\right)_\mathrm{ini},
\end{equation}
where the ratio $\delta \varphi/\varphi$ behaves like a constant when
$V(\varphi)$ is quadratic.
Otherwise, the result is highly model-dependent.
Note that the deviation from $\alpha\sim 1$ might be significant at the
waterfall if the evolution is already far from slow-roll (i.e, when
$\beta\gg 1$).
In this case, the non-linear evolution of $\varphi$ and $\delta \varphi$
is significant and it requires further study. 
Here, considering a model with the modest evolution $\alpha\sim 1$ and
$\beta\sim 1$ (as motivated by supergravity), 
we expect $r_\mathrm{osc}\sim 0.1$.

\section{Inflating Hybrid curvaton}

It is possible to realize the inflating curvaton scenario 
if the oscillation starts after curvaton domination.
This scenario always gives $r_\mathrm{osc}=1$ by definition, and helps to
explain the 
spectral index and its running by reducing the number of e-foldings
required for 
the primordial inflation~\cite{infla-curv}.
The number of e-foldings during the curvaton inflation ($N_2$) is
determined by the initial condition $\varphi(t_\mathrm{ini})$ and $\beta$;
\begin{equation}
\label{N2}
N_2 =\frac{1}{\beta}\ln \frac{\varphi(t_\mathrm{ini})}{\varphi_c},
\end{equation}
where $t_\mathrm{ini}$ denotes the beginning of the curvaton inflation.
In this section we assume ${\cal O}(1)$ couplings for the hybrid
potential ($\lambda\simeq g\simeq 1$) just for simplicity.
Considering the bound $\varphi(t_\mathrm{ini})<M_p$ and $M > 1$TeV
(i.e, $\varphi_c >1$TeV),
we find from Eq.~(\ref{N2}) that the maximum number of $N_2$ is given by
\begin{equation}
\label{35beta}
N_2 <35\beta^{-1}.
\end{equation}
Slow-roll ($\beta \ll 1$) requires $\varphi(t_\mathrm{ini}) \sim
\varphi_c$ if we demand  $N_2 < 60$.
We are not excluding this possibility, but it would be interesting to
relax this tuning using a fast-roll potential ($\eta\sim \beta\sim 1$).
In this case we find
\begin{equation}
\delta N_2 =\frac{1}{\beta}\frac{\delta \varphi}{\varphi}
\sim \frac{\delta \varphi}{\varphi}.
\end{equation}
Since the ratio $\delta \varphi/\varphi$ does not evolve in the case of a
quadratic potential, we obtain
\begin{equation}
\label{jyougen}
{\cal P}^{1/2}_{\zeta}\sim
\frac{H_1}{2\pi \varphi_c e^{\beta N_2}},
\end{equation}
where the relation $\varphi(t_\mathrm{ini}) \simeq \varphi_c e^{\beta
N_2}$ was used. 
Here $ {\cal P}^{1/2}_{\zeta}$ denotes the observable spectrum from
the CMB. 

If we consider the stochastic initial
condition \mbox{$H_1>M$} \cite{infla-curv}, we find 
${\cal P}^{1/2}_{\zeta}  \gg e^{-\beta N_2}$.
Here  $\varphi_c\simeq M$ is used because 
${\cal O}(1)$ couplings are assumed.
Therefore, the stochastic condition leads to
\begin{equation}
N_2 > 13\beta^{-1} \sim 13.
\end{equation}
This result is in sharp contrast to the PNGB inflating
curvaton~\cite{infla-curv}, in which the stochastic condition puts an
{\em upper} bound for $N_2$.

An interesting application of this scenario is that $M$
can be related to the dynamical scale of supersymmetry breaking
while $\beta\sim 1$ is due to the supergravity action.
Considering $M\sim \Lambda_{s}\sim 10^{11}$GeV, we find
(The upper bound is different from Eq.(\ref{35beta}) because the scale
of $M$ is different.) 
\begin{equation}
13< N_2 <16,\,\,\,\,\, (10^{16} {\rm GeV}  <\varphi(t_\mathrm{ini})<M_p)
\end{equation}
and for  $M\sim 10^{6}$GeV we find
\begin{equation}
13< N_2 < 27, \,\,\,\,\, (10^{11}{\rm GeV}
 <\varphi(t_\mathrm{ini})<M_p).
\end{equation}

As in the standard curvaton scenario, the spectral index is given by
\begin{equation}
n-1=-2\epsilon_1+2\eta,
\end{equation}
where $\epsilon_1$ (for the primordial inflaton) and $\eta$ (for the
slow-roll curvaton field) are calculated at the horizon exit.
If the primordial inflation is of the chaotic type with potential
$V(\phi)\simeq A\phi^p$, then $\epsilon_1\simeq p/4N_1$.
According to observations, $n-1=-0.037\pm 0.014$~\cite{wmap7}, which leads to
$N_1 \simeq 14p$ for negligible $\eta$.
In the oscillating curvaton scenario ($N_2=0$) the spectral index 
excludes $p<4$, and gives for the running of the spectral index
$n'\equiv (n-1)/N_1\simeq 0.0007$.
The situation is quite different in the inflating curvaton scenario.
Because of the second inflation $N_2\ne 0$, $p<4$ is not ruled out,
while the running of the spectral index is enhanced when 
$N_1<60$, which can have observational impact~\cite{infla-curv}.

\section{Conclusions}

One of the advantages of the curvaton idea is that the sector, which
gives rise to primordial inflation (inflaton sector), can be decoupled
from the sector which gives rise to the curvature perturbation (curvaton
sector). 
This can naturally accommodate the stability of the hierarchy in
scales, which is a feature of the curvaton scenario. 
Moreover, in the
curvaton scenario, the inflaton sector does not even need to feature a
fundamental scalar field at all~\footnote{The use of $f(R)$ gravity in
the inflationary cosmology was pioneered by
Starobinsky~\cite{Starobinsky:1980te}.}
 In that sense, the curvaton scenario liberates the inflaton sector from
 many serious constraints~\cite{Dimo-Lyth}.

Although the inflaton sector has been investigated in the context of
various theoretical models, including the hybrid-type potential for the
inflaton field, the curvaton sector is usually considered merely with
the simple quadratic potential. 
This gives somewhat ``biased'' information about the possibilities for
inflationary cosmology.  
In that sense, our paper is an effort to expand the state of the
art. 
Furthermore, the hybrid potential, which is the target in this
paper, is well motivated in particle physics without being necessarily
tied to the inflaton. 
For example, the hybrid structure may appear in
GUTs or brane-brane interactions~\cite{brane-angled,
GarciaBellido:2000gs} while it can well explain the differences in
scales~\cite{GarciaBellido:2000gs}. 
Finally, in the hybrid curvaton, the source mechanism of the curvature
perturbation is utterly different from the original curvaton
scenario. 
In that sense the basic formulation is brand-new.

To be more specific, we considered in this paper a new possibility of
the curvaton scenario; we studied oscillating, inflating and non-slow-roll
scenarios for the hybrid curvaton.  For the slow-roll scenarios
we found significant enhancement of the curvature perturbation, while
in the non-slow-roll scenarios we found that the tuning requirements for
the initial condition are relaxed. The inflating curvaton predicts 
distinguishable running of
the spectral index, which cannot be attained in the usual inflation
scenario except for a few extended models.\footnote{ Some extended
models can attain significant running even for single-field inflaton
models. for instance, the running mass model~\cite{runnin-mass} and the
Type III Hilltop inflation model~\cite{hilltop3} predict such a large
running of the spectral index.}  There is a variety of models possible
for the hybrid curvaton.  For example, one can consider the
curvaton analogues of smooth hybrid  inflation~\cite{smooth} or
shifted hybrid inflation~\cite{shifted}.

Since the purpose of this paper is not the model building but the basic
formulation of the brand-new idea, we avoided highly model-dependent
arguments and focused on the simplest model that
is useful to explain the essential idea of the scenario. 
We have considered $\alpha=3(H\dot\varphi/V_\varphi)_{\rm osc}
\sim 1$ because it is the natural choice. However, the scenario with 
$\alpha \gg 1$ is interesting as well.
A large $\alpha$ would reduce the spectrum of the curvature
perturbation, but the non-linear evolution might enhance
non-Gaussianity, in a way that has not been considered yet. 
However, the calculation requires numerical study.
We have also considered $\beta\sim 1$ ($\beta$ is defined in
Eq.~(\ref{beta})) because this is well motivated by supergravity.
$\beta$ is generically a model-dependent parameter and,
were it not for the supergravity motivation, there would be no reason to 
believe that $\beta\sim 1$.
Finally, we considered that the oscillation starts immediately after the
end of the slow-roll evolution to avoid the complexity related to the dynamics 
of the waterfall. However, in some cases this
dynamics could be non-trivial, which may
require numerical calculations. Indeed, the
evolution of the waterfall is not easily understood by analytical calculation
only~\cite{Lyth-hybrid} even without a radiation background.

\section{Acknowledgment}
We thank David H. Lyth for valuable discussions.
KD is supported (in part) by the Lancaster-Manchester-Sheffield 
Consortium for Fundamental Physics under STFC grant ST/J000418/1.
KK is partly supported by the Grant-in-Aid for the Ministry of  
Education, Culture, Sports, Science and Technology, Government of
Japan Nos. 21111006, 22244030, 23540327, and by the Center for the 
Promotion of Integrated Science (CPIS) of Sokendai (1HB5806020).

\end{document}